**Optimization of an underwater in-situ $LaBr_3$:Ce spectrometer with energy self-calibration and efficiency calibration**

Zhi Zeng[1], Xingyu Pan[1], Hao Ma[1], Jianhua He[2,3], Jirong Cang[1], Ming Zeng[1*], Yuhao Mi[1] Jianping Cheng[1]

[1] Key Laboratory of Particle and Radiation Imaging (Ministry of Education) and Department of Engineering Physics, Tsinghua University, Beijing 100084, China

[2] Laboratory of Marine Isotopic Technology and Environmental Risk Assessment, State Oceanic Administration, Xiamen 361005, China;

[3] Third Institution of Oceanography, State Oceanic Administration, Xiamen 361005, China.

## Highlights

- An underwater in-situ $LaBr_3$:Ce spectrometer was developed and tested.
- The intrinsic background of $LaBr_3$:Ce was well determined by underground $\gamma$ spectrometer and $\alpha$ / $\gamma$ discrimination.
- A self-calibration method including spectra stabilization was proposed and validated.
- The efficiency was calculated by water tank experiment and MC simulation.

## Abstract

An underwater in-situ gamma-ray spectrometer based on $LaBr_3$:Ce was developed and optimized to monitor marine radioactivity. The intrinsic background mainly from $^{138}$La and $^{227}$Ac of $LaBr_3$:Ce was well determined by low background measurement and pulse shape discrimination method. A method of self-calibration using three internal contaminant peaks was proposed to eliminate the peak shift during long-term monitoring. With experiments under different temperatures, the method was proved to be helpful for maintaining long-term stability. To monitor the marine radioactivity, the spectrometer's efficiency was calculated via water tank experiment as well as Monte Carlo simulation.

## Key words

Underwater $\gamma$-ray spectrometer; $LaBr_3$(Ce); temperature effect; self-calibration; MC simulation; efficiency calibration

## 1. Introduction

Underwater in-situ gamma spectrometers with NaI(Tl) detector were wildly used to monitor the marine radioactivity continuously (Casanovas et al., 2013, Thornton et al., 2013, Tsabaris et al.,

*Corresponding author: Ming Zeng, email address: zengming@tsinghua.edu.cn.

2008, Osvath et al., 2005). To obtain better energy resolutions, such spectrometers with HPGe (Provinec et al., 1996) or $LaBr_3$:Ce detector (Su et al., 2011) had also been applied. Especially, the γ-ray spectrometers with $LaBr_3$:Ce detector perform better than those with NaI(Tl) detector (Zeng et al., 2015), including better energy resolution (2.7%@662keV), high light yield (63 photons/keV) and good energy linearity. However, the intrinsic radioactivity background in the $LaBr_3$:Ce crystal from $^{138}$La and $^{227}$Ac decay chain would make the MDA(minimum detectable Activity) worse if not removed properly(Su et al., 2011, Casanovas et al., 2014).

Considering the energy resolution and calculation of activity, there are two main problems which should be solved when applying in-situ gamma spectrometer in marine monitoring:

1) The peak-shift caused by temperature variance and other electronic factors;

2) The efficiency calibration, which can be determined either experimentally or using MC simulations.

For in-situ spectrometers with $LaBr_3$:Ce detector, the gamma peak caused by the intrinsic radioactivity can be used as an indicator to eliminate the temperature effect and improve the system's long-term stability (Moszyński et al., 2006). For example, a temperature peak-shift correction method using a single known peak in $LaBr_3$:Ce intrinsic gamma spectrum to realize energy self-calibration has been proposed (Casanovas et al., 2012). Also, a second-order polynomial relationship between the peak-shift and corresponding temperature has been established to correct the temperature peak-shift with several standard gamma sources. In general, these methods are good enough below 1.7MeV since the relative channel displacement, i.e. channel displacement in certain channel divided by its initial position of the channel, is approximately the same for the region of 0~1.7MeV in spectra. However, in higher energy region of 1.7~3MeV, slight difference of relative channel displacement can be distinguished which may result in a deviation of about 7.5~15keV with a single known peak. In this situation, using two or more known peaks may provide a better estimation of the peak shift in all channels, especially if the chosen peaks cover a broad energy range in spectra.

To calculate activity of interested radionuclide, the detection efficiency of spectrometers at different energies needs to be estimated. There are mainly two methods: a) Monte Carlo simulation (Bagatelas et al., 2010, Zeng et al., 2015); b) measuring diluted standard sources in water tank, including $^{99m}$Tc(140.5keV, $T_{1/2}$~ 6 hours), $^{137}$Cs(661.7keV, $T_{1/2}$ ~30.17y) and $^{40}$K(1460.8keV, $T_{1/2}$ ~$1.27\times10^9$ y) (Tsabaris et al., 2008). A big difference is found between the efficiency of lower energy (below 150~200keV) and higher energy (over 200~300keV) for $LaBr_3$:Ce detector. And the half-life of $^{99m}$Tc is too short to calibrate gamma spectrometer in a long time. As a result, if a radionuclide with long half-life and low-energy gamma line can be used, the efficiency in low energy can be more accurate.

In this work, a peak-shift correction method with three known peaks from intrinsic radioactivity background in $LaBr_3$:Ce detector was proposed to improve long-term stability. And, the detection efficiency was calculated with MC simulation and water tank experiments. To maintain more accurate efficiency, $^{238}U$-$^{234}Th$ decay chain, in which a 92.5keV γ-ray was emitted, was chosen as the lower energy calibration source. To validate the methods, two γ-ray spectrometers, equipped with a 3"×3" $LaBr_3$:Ce crystal and a 2"×2" one (BrilLance™ 380, Saint-Gobain) respectively, were built and tested.

## 2. Materials and methods

### 2.1 Experimental set-up

The γ-ray spectrometers used in this study were equipped with a 2"×2" and a 3"×3" $LaBr_3$:Ce crystals from Saint-Gobain (BrilLance™ 380). Both crystals were connected to a photomultiplier tube, high voltage supplier, Amptek DP5 multichannel analyzer and etc. A temperature sensor in DP5 board was used to reflect temperature of the module, which is also an indicator of the temperature of scintillator. The number of channels in the MCA was 2048 and the energy threshold was set to be 25keV. More information about the detailed structure has been described in a previous article (Zeng et al., 2015).

To determinate the radionuclides inside $LaBr_3$:Ce detector, an underground low background gamma spectrometer called GeTHU in China JinPing underground Laboratory(CJPL) was used (Zeng et al., 2014). CJPL is the deepest underground laboratory whose rock overburden is about 2400 meters, and the cosmic ray flux in CJPL is only $2\times10^{-10}/cm^2/s$ (Wu et al., 2013). GeTHU is based on a coaxial n-type HPGe detector of 40% relative detection efficiency. And it is shielded with 10 cm polyethylene, 15 cm lead and 5 cm copper from outside to inside, which contributes to an excellent detection sensitivity.

To imitate the actual underwater environment, two water tanks full of water added with artificial radioactive nuclides were prepared. Both tanks are $\Phi2.0m\times2.3m$, which are large enough according to the effective detection distance of about 80cm for 3"×3" $LaBr_3$:Ce detector (Zeng et al., 2015). One tank was added with $0.46\pm0.01Bq/L$ $^{137}Cs$ and the other one was added with $U_3O_8$ which contributes to $5.86\pm0.69Bq/L$ $^{238}U$ in the tank. As is shown in Fig 1, the half-life of $^{234}Th$ and $^{234m}Pa$ is quite small while the half-life of $^{234}U$ is very long, so the secular equilibrium ends at $^{234m}Pa$ for pure $U_3O_8$ material. Two γ-rays from $^{234}Th$, the 92.38keV (2.18%) one and the 92.80keV (2.15%) one with the biggest emission probability, can be treated as an indicator of the activity of $^{238}U$.

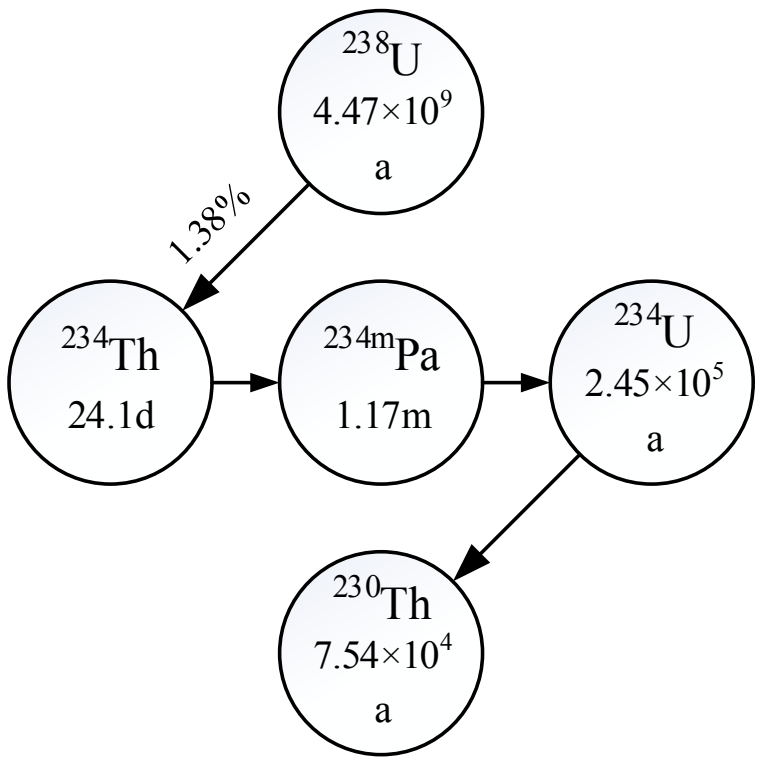

Fig 1 Radioactive series of $^{238}U$.

## 2.2 Intrinsic background of $LaBr_3$:Ce

The intrinsic radioactivity of $LaBr_3$:Ce mainly comes from $^{138}La$ (0.09% naturally abundant) and $^{227}Ac$ (actinium is a chemical analogue to lanthanum). The decay of $^{138}La$ has two modes. One (65.2%) is electron capture to the excited level of $^{138}Ba$ with ensuing emission of de-excitation γ-ray (1435.8keV). Besides, characteristic X-rays, originate from the cascade radiation following the electron capture in the K- and L-shell of $^{138}Ba$, are emitted as well. Since the cascade occurs inside the crystal itself, the binding energy of K and L shell of $^{138}Ba$ should be added with the γ-ray energy and produces two peaks at 1472keV and 1441keV, i.e. the sum peak of 1436keV γ-ray with 5keV L-shell cascade radiation and 36keV K-shell cascade radiation. The other decay mode is a β- decay to the excited level of $^{138}Ba$ with consequent γ-ray (789keV), which produces a β continuum between 789keV and ~1044keV. The other origin of intrinsic background is $^{227}Ac$, one of nuclides in actinium series, which produces a series of radionuclides (mainly α-decay nuclides) in the crystal. In spectra, the contamination of $^{227}Ac$ and its daughters contributes to the background in the region of 1.6~3MeV.

## 2.3 Energy self-calibration algorithm

The energy calibration of a spectrometer is constant when the monitoring condition is stable, which means the relationship between channel and deposited energy is constant. However, the temperature dependency of light yield of $LaBr_3$:Ce is a major obstacle of the long-term stability. Furthermore, the dark current of photocathode, the secondary electron emission yield of dynode and some other electronic components, such as amplifier, power supply, can be affected by the change of temperature. The direct effect of temperature change is peak shift, which influences the search of photopeak and the calculation of the activity of corresponding radionuclide. However, for long-term underwater monitoring, it is not feasible to calibrate spectrometer with standard sources when temperature changes. To solve the problem, the intrinsic radioactivity of $LaBe_3$:Ce could be applied for self-calibration.

First of all, the spectrometer should be calibrated under a certain condition. Thus, a calibration curve of the detector is established under the corresponding condition $S_0$. The calibrated spectrum is treated as a standard spectrum applied during the later self-calibration procedure. To eliminate the nonlinearity of scintillator, a second-order polynomial equation between the measured $i^{th}$-channel position $C_i^0$ and deposited energy $E_i^0$ under the condition $S_0$ is established as follow:

$$E_i^0 = a_0 + a_1 \cdot C_i^0 + a_2 \cdot (C_i^0)^2 \tag{1}$$

As is illustrated before, the intrinsic radioactivity from $^{138}$La and $^{227}$Ac decay will cause several extra peaks, which are chosen to maintain the self-calibration. When monitoring underwater environment under the condition $S_1$, channels of the intrinsic background peaks are recorded as $\{C_{p1}^1, C_{p2}^1, \cdots\}$. And the channels of these peaks in standard spectrum are recorded as $\{C_{p1}^0, C_{p2}^0, \cdots\}$. A second-order polynomial relationship between $\{C_{p1}^1, C_{p2}^1, \cdots\}$ and $\{C_{p1}^0, C_{p2}^0, \cdots\}$ is established:

$$C_{pi}^0 = b_0 + b_1 \cdot C_{pi}^1 + b_2 \cdot (C_{pi}^1)^2 \tag{2}$$

For the measured $i^{th}$-channel position $C_i^1$ under condition $S_1$, the peak-shift can be corrected utilizing equation (2):

$$C_i^{corrected} = b_0 + b_1 \cdot C_i^1 + b_2 \cdot (C_i^1)^2 \tag{3}$$

Applying the corrected channel in equation (1), corresponding deposited energy can be obtained:

$$E_i^1 = a_0 + a_1 \cdot C_i^{corrected} + a_2 \cdot (C_i^{corrected})^2 \tag{4}$$

Combining equation (3) and (4), all the channels in measured spectrum can be corrected once those peaks are determined, thus the self-calibration can be realized accurately to eliminate the peak-shift.

## 2.4 Methods of efficiency calibration

The efficiency of an underwater in-situ $LaBr_3$:Ce spectrometer is defined as $\varepsilon_V = \dot{n}_0/A_V$, where $\varepsilon_V$ represents the volumetric efficiency in cps/(Bq/L), $\dot{n}_0$ stands for the photopeak count rate, and $A_V$ represents activity concentration in Bq/L in marine (Su et al., 2011). Assuming that one kind of radionuclide with concentration of A(Bq) is distributed uniformly in a tank with volume of V(L) and its γ-ray emission probability is $I_\gamma$, the photopeak counting rate can be calculated by:

$$\dot{n}_0 = A \cdot I_\gamma \cdot \varepsilon_{sp} \tag{5}$$

where $\varepsilon_{sp}$ means the ratio of detected γ-rays to emitted γ-rays. So the detection efficiency can be calculated:

$$\varepsilon_V = \dot{n}_0/A_V = I_\gamma \cdot \varepsilon_{sp} \cdot V \tag{6}$$

# 3. Results and discussion

## 3.1 Intrinsic background measurement at CJPL

### 3.1.1 Intrinsic gamma radioactivity of $LaBr_3$:Ce

To determination the gamma radioactivity of $LaBr_3$:Ce crystal, γ spectrometer with high sensitivity can be a good candidate. The measurement using a HPGe detector was applied already (Camp, et al., 2016) and the contribution of $^{138}$La and $^{40}$K was calculated, however the γ-rays emitted from daughter nuclides of $^{227}$Ac was not recognized in the spectrum. To recognize and quantitate the contributions from different gamma radionuclides in $LaBr_3$:Ce, a commercial $LaBr_3$:Ce crystal (BrilLance™ 380, Saint-Gobain) was well measured with GeTHU, an underground low background gamma spectrometer in CJPL. In this work, a 2"×2" $LaBr_3$:Ce crystal was measured with GeTHU. The measured parts include a 2"×2" $LaBr_3$:Ce crystal covered by an aluminum housing, a photomultiplier, a voltage divider and preamplifier module, as is shown in Fig 2.

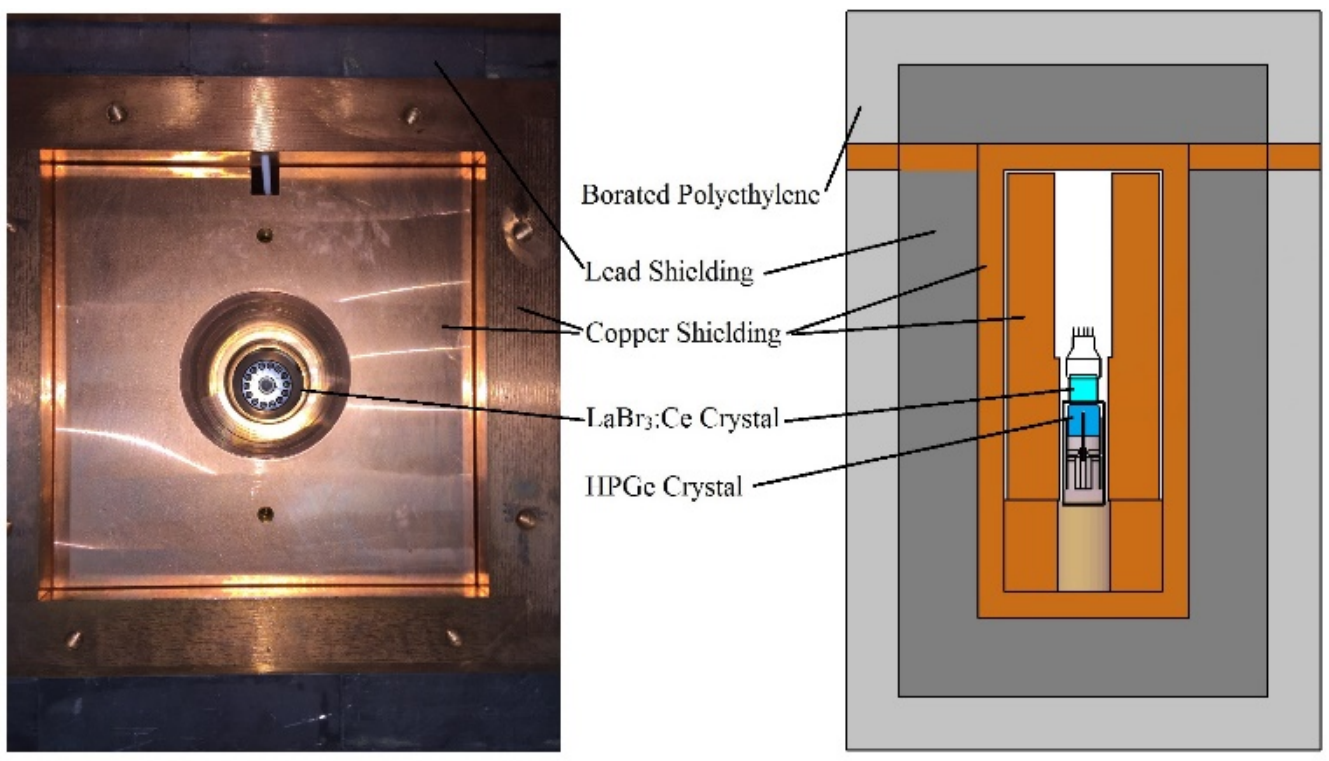

Fig 2 Measurement scheme (top view, left) and simulation geometry (right) of 2"×2" $LaBr_3$:Ce crystal in GeTHU.

As Fig 3 showed, the main radioactivity is contributed by $^{138}$La and $^{40}$K. Besides, several gamma peaks from daughter nuclides of $^{227}$Ac including $^{227}$Th, $^{223}$Ra, $^{219}$Rn, $^{211}$Bi were found as well. To quantitate the activities of those radionuclides, GEANT4 code was used to simulate the detection efficiency of GeTHU with real geometry of measurement, as is shown in Fig 2(right). With simulated detection efficiency, the activities of intrinsic radionuclides can be calculated, shown in Table 1. The activity of $^{138}$La is calculated to be 133±5Bq wtih the photopeaks in 789keV and 1436keV emitted by $^{138}$La. On the other hand, according to the natural abundance of $^{138}$La, the activity of $^{138}$La in $LaBr_3$:5%Ce was also calculated theoretically, whose result is about 151.0Bq (Rosson et al., 2011). The deviation of $^{138}$La activity between γ analysis and theoretical calculation is maybe caused by the difference in geometry setup and materials. This is the main uncertainty since the geometry and materials used in Monte Carlo simulation were refer to the x-ray imaging of 2"×2" $LaBr_3$:Ce detector in this work. The activities of $^{40}$K, $^{227}$Th and $^{211}$Bi were also calculated with the simulated efficiencies as showed in Table.1. As a result, $^{227}$Ac and its daughter nuclides, $^{227}$Th and $^{211}$Bi , are in secular equilibrium in the crystal.

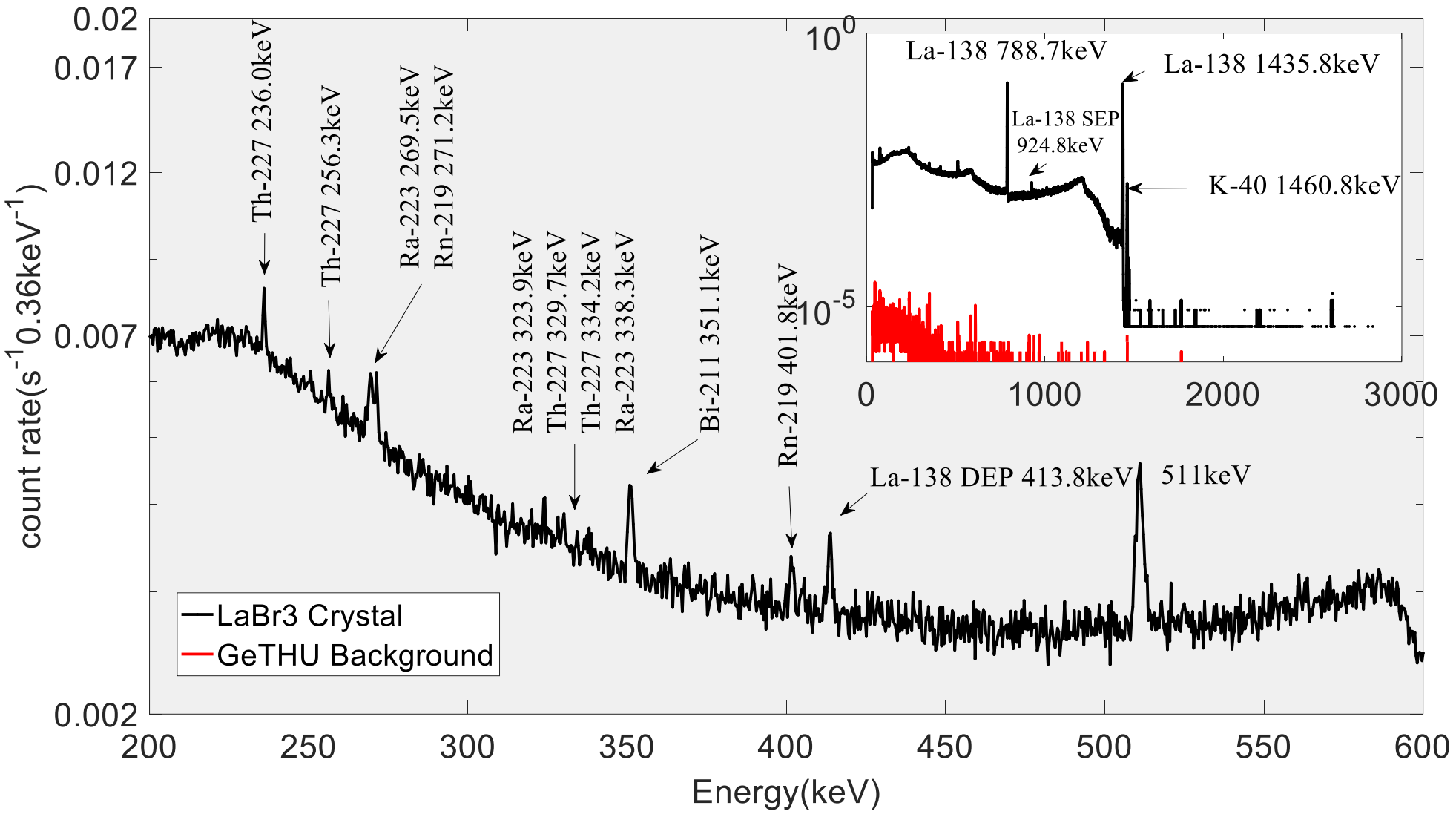

Fig 3 The gamma spectrum of 2"×2" $LaBr_3$:Ce crystal measured by GeTHU. Count rates are in counts per second and per channel of the MCA in GeTHU, which represents 0.36keV in energy.

Table 1 Intrinsic radioactivity of 2"×2" $LaBr_3$:Ce crystal measured by GeTHU

| Nuclides | $^{138}$La | | $^{40}$K | $^{227}$Th | $^{211}$Bi |
|---|---|---|---|---|---|
| Gamma energy(keV) | 788.7 | 1435.8 | 1460.8 | 236.0 | 351.1 |
| Efficiency(cps/Bq) | 0.0118 | 0.0084 | 0.0091 | 0.0184 | 0.0186 |
| Intensity(%) | 34.8 | 65.2 | 10.55 | 11.50 | 12.95 |
| Counting rate(cps) | 0.560 | 0.712 | 0.0114 | 0.0044 | 0.0054 |
| Activity(Bq) | 136.1±7.3[a] | 130.5±7.0[a] | 11.8±0.7[b] | 2.1±0.1 | 2.2±0.1 |

*a: The theoretical activity of $^{138}$La in 2"×2" $LaBr_3$:5%Ce crystal is about 151.0Bq.*

*b: The $^{40}$K maybe exist in PMT since the whole detector include crystal and PMT are measured together.*

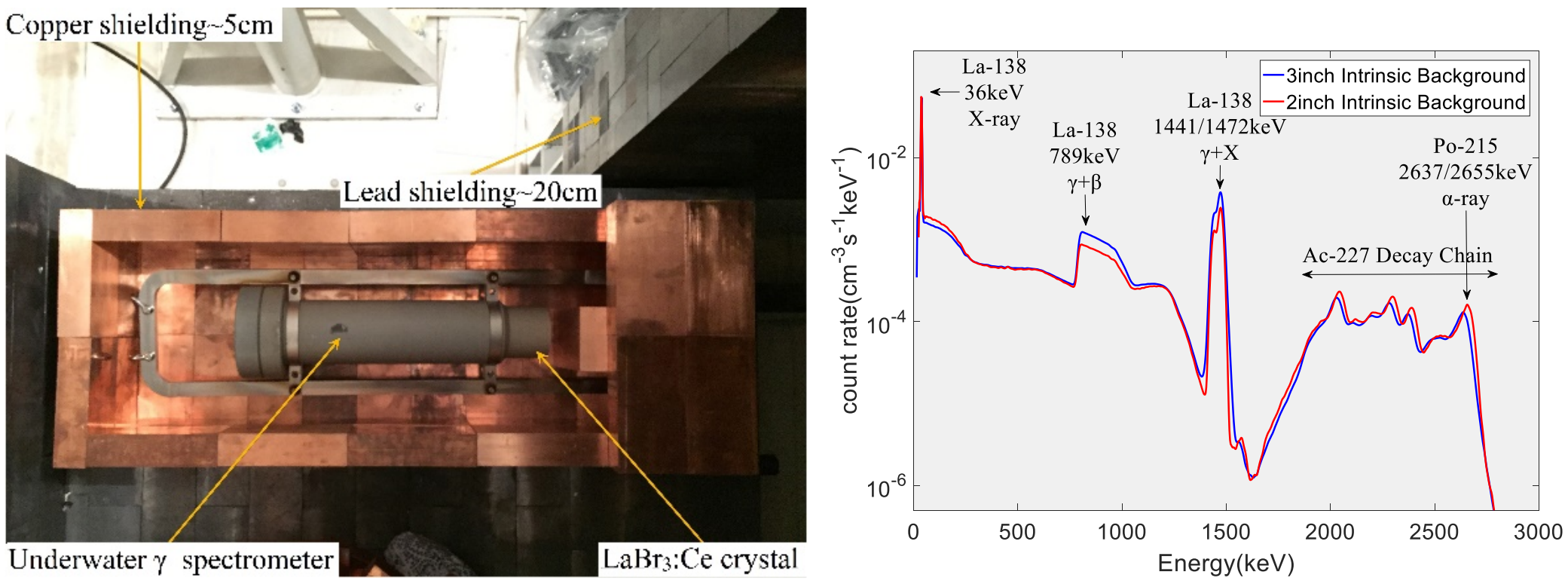

Fig 4 The shielding device for intrinsic background spectra measurement in CJPL(left) and the intrinsic background spectra of 2"×2" and 3"×3" $LaBr_3$:Ce detectors(right). Count rates are in counts per $cm^3$ (volume of crystal) per second and per keV.

Except the measurement with HPGe detector, the intrinsic background spectra of 2"×2" and 3"×3" $LaBr_3$:Ce detectors were measured under well-shielding environment at CJPL as well. The $LaBr_3$:Ce detectors were put into a heavy shielding device inside a polyethylene (PE) room surrounded by 100 cm thick polyethylene in CJPL. The inner shield surrounding the detector is 5 cm thick copper, and the outer is 20 cm thick lead brick, shown in Fig 4(left). Before the measurement, the two detectors were calibrated with standard sources including $^{241}$Am、$^{137}$Cs and $^{60}$Co, the calibration curves were fitted with second-order polynomial equations as follows:

$$\begin{aligned} Energy(3inch) &= 2.49\times10^{-5}Channel^2 + 1.32\times Channel + 0.71 \\ Energy(2inch) &= 1.97\times10^{-5}Channel^2 + 1.42\times Channel - 0.0079 \end{aligned} \quad (7)$$

With the equations, the measured spectra of 2"×2" and 3"×3" $LaBr_3$:Ce detectors were calibrated, shown in Fig 4(right). In the spectra, the 36keV peak originates from the cascade radiation following the electron capture in the K-shell of $^{138}$Ba. With a proper threshold of the detector, it can be recognized clearly. In the region of 1400~1500keV, two peaks can be recognized by double-Gaussian fitting with the central energy of 1441keV and 1472keV.

### 3.1.2 Intrinsic alpha radioactivity of $LaBr_3$:Ce

There is a continuum peak lying in the background spectra of 2"×2" and 3"×3" $LaBr_3$:Ce detectors between 1.6~3.0MeV. In our previous work, an α/γ discrimination method based on pulse shape discrimination for 2"×2" $LaBr_3$:Ce detector was applied to study this area carefully (Zeng et al., 2016). A 2.5 Gsps 12-bit LeCroy Oscilloscope (HDO6104) was used to digitize the raw PMT output at common indoor environment without any shielding devices. As shown in Fig 5, this continuum can be divided into two parts:

1) α spectrum (the blue line), similar with the background;

2) γ spectrum (the brown line), emitted by $^{208}$Tl(2615keV) in the environment.

Thus, it can be concluded that the counts in 1.6~3MeV mainly owe to the α-rays emitted by $^{227}$Ac and its daughter nuclides. Furthermore, compared with the intrinsic background measured by the detector itself in CJPL (the black line), an obvious lack of counts near 2650keV in the spectrum collected by oscilloscope (the red line) can be found because the half-life of $^{215}$Po is 1.781ms, which is much less than the dead time of the oscilloscope (~50 ms). Thus, the peak with highest energy is believed to be the contribution of the 7386keV α particle from $^{215}$Po, the daughter radionuclide of $^{227}$Ac. As is shown in Fig 4(right), the gamma equivalent energies of α particles of 2"×2" and 3"×3" $LaBr_3$:Ce detectors are different. For example, the equivalent energy of $^{215}$Po in 2"×2" $LaBr_3$:Ce detector is 2655keV, while in 3"×3" detector it is 2637keV. The difference of equivalent energies is more serious in co-doped crystals, where the equivalent energy of $^{215}$Po is about 3.5MeV (Yang et al., 2016). Thus, the difference is possibly due to the slight difference of compositions in the two crystals.

As a result, the intrinsic background of $LaBr_3$:Ce can be divided into two parts: the region of 25keV~1.6MeV and 1.6~3.0MeV, coming from $^{138}$La and $^{227}$Ac decay chain respectively, and the integral count rates in each region are 1.27cps/cm$^3$ and 0.094cps/cm$^3$ for 2"×2" $LaBr_3$:Ce. As for 3"×3" $LaBr_3$:Ce, the integral count rates are 1.32cps/cm$^3$ and 0.082cps/cm$^3$, which is in good agreement with the result in other $LaBr_3$:Ce crystals (Quarati et al., 2012).

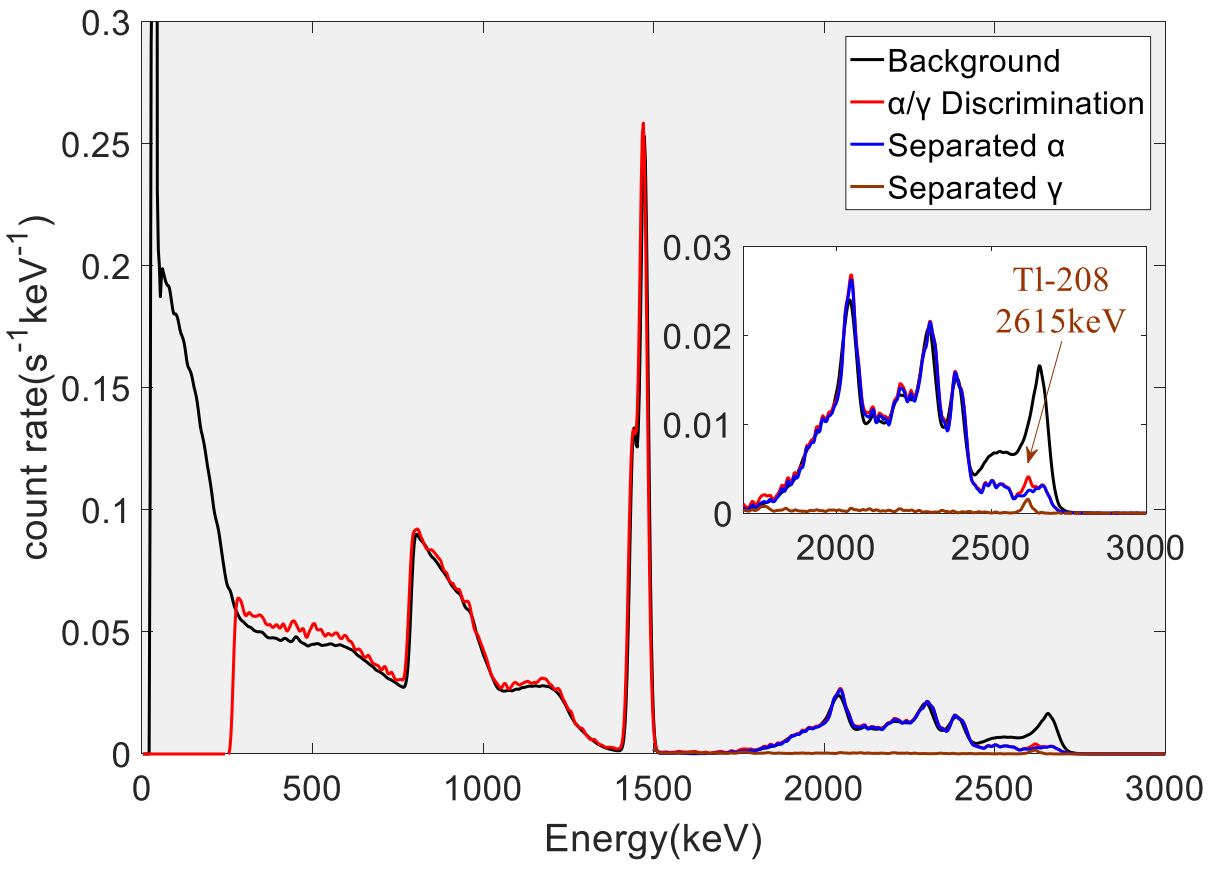

Fig 5 α/γ discrimination for 2"×2" $LaBr_3$:Ce detector.

### 3.2 Validation of self-calibration algorithm

#### 3.2.1 Temperature dependency of peak shift

To have an overall understanding of the relationship between temperature and peak shift, a series of experiments with 3"×3" $LaBr_3$:Ce detector were carried out in the water tank mentioned before. The experiments lasted for over 1400 hours and measured spectra and corresponding temperature were stored every one hour.

Three raw measured spectra under different temperatures were shown in Fig 6(a) to reflect the temperature effect of $LaBr_3$:Ce detector. To have a better demonstration of temperature effect, the channel positions of 7 distinguishable peaks in the region of 0~3.0MeV were extracted, including the sum peak of 789keV γ-ray and β continuum. To eliminate the difference caused by energies of different peaks, relative channel displacement (Casanovas et al., 2012) was used to describe floating channels of peaks with different energies, which can be calculated with the following equation:

$$R_i(\%) = (C_{pi}^1 - C_{pi}^0)/C_{pi}^0 \cdot 100 \quad (8)$$

where $R_i$ represents the relative channel displacement in $i^{th}$-peak, $C_{pi}^1$ stands for the channel of $i^{th}$-peak in spectrum measured under the condition $S_1$, and $C_{pi}^0$ represents the channel of $i^{th}$-peak in a reference spectrum measured under the condition $S_0$. The relative channel displacements of 7 peaks under different temperatures were shown in Fig 6(b). In the region below 1.5MeV, the relative channel displacement is almost the same, which means the method of correcting the shifted channel with a proportional relationship using a single peak, as is mentioned previously (Casanovas et al., 2012), is efficient. However, in the region over 1.5MeV, the relative channel displacement seems to be different and in some case the difference could reach to about 0.5%, which means a deviation of about 7.5~15keV may occur in this region if the spectrum is calibrated using a single peak. Thus, to have a better calibration for the whole energy range, a second-order polynomial equation between channels in actual and standard spectrum is adopted, as is described before in section 2.3.

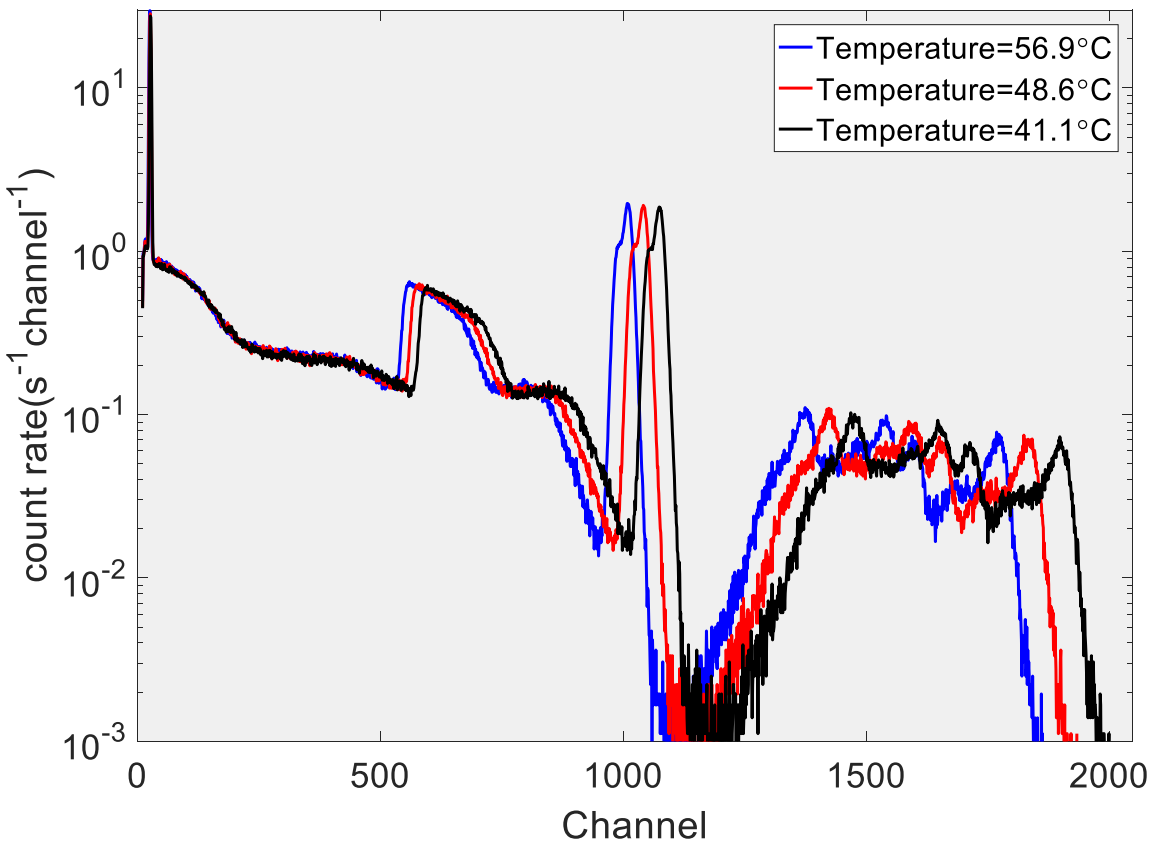

(a) Raw measured spectra

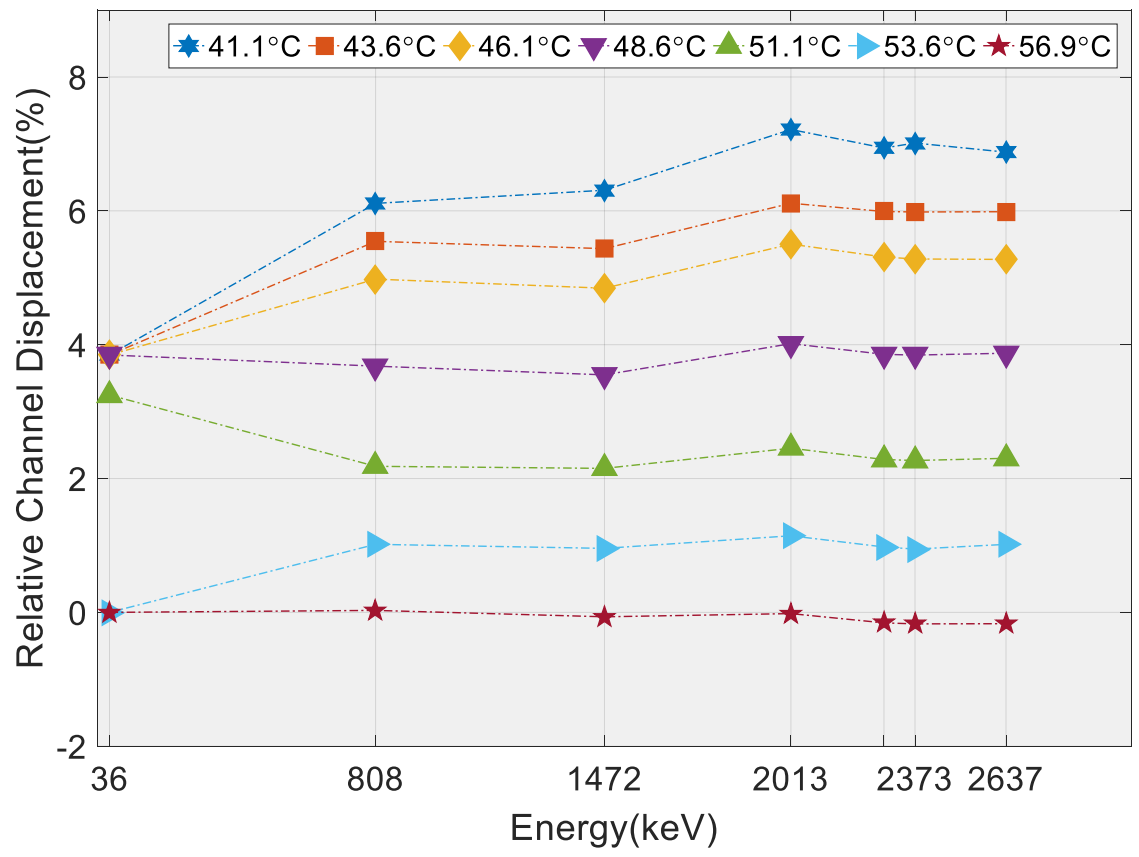

(b) Temperature dependency of peak shift

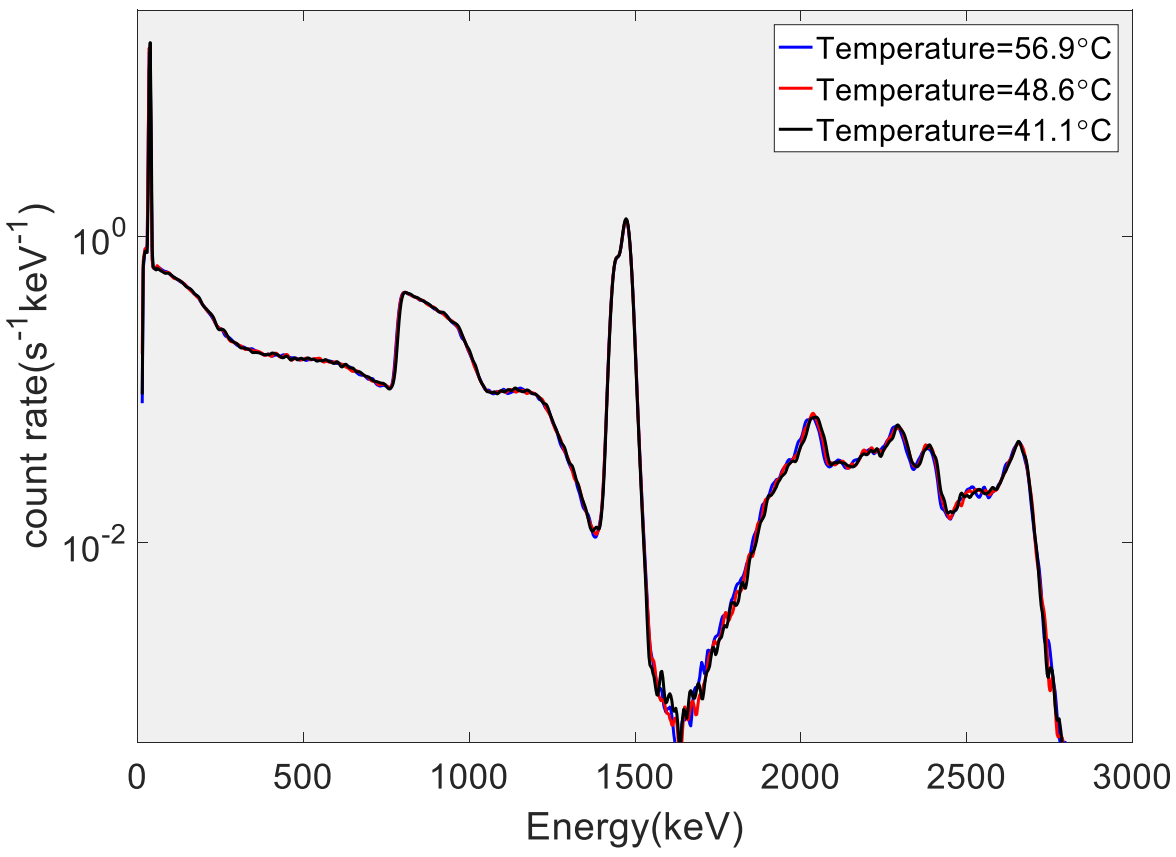

(c) Corrected spectra (calibrated)

Fig 6 Temperature effect and self-calibration method.

### 3.2.2 self-calibration energy peak candidates of $LaBr_3$:Ce

To establish a second-order polynomial relationship between channel of standard spectrum and measured spectrum, at least three known peaks is needed. In the intrinsic background spectra of $LaBr_3$:Ce, as is shown in Fig 4(right), several peaks of different radiation types including X, γ and α rays can be distinguished apparently, which provides an access to correct the peak-shift caused by temperature or in other words to maintain self-calibration. To eliminate the nonlinearity, the chosen peaks for self-calibration should cover the whole energy region of 0~3MeV. Furthermore, the chosen peaks should not overlap γ peaks of monitoring radionuclides and the counts of chosen peaks should be high enough to be distinguished within a limited time (10 minutes or so). Considering the above demands, three peaks were chosen to maintain self-calibration:

a) The 36keV peak: the K-shell cascade radiation from $^{138}$Ba;

b) The 1472keV peak: the sum peak of 1436keV γ-ray and 36keV X-ray from $^{138}$La;

c) The 2637/2655keV peak: the α peak from $^{215}$Po.

### 3.2.3 Validation of self-calibration algorithm

Three peaks caused by the internal contaminant of the $LaBr_3$:Ce crystal were used to calibrate the detector. Remarkable results of the self-calibration for 3"×3" $LaBr_3$:Ce detector are shown in Fig 6(c), which eliminates the peak shift during long-term monitoring effectively. The usage of three peaks, which cover the whole energy region of 0~3MeV, helps a lot to remove the effect of nonlinearity. The deviation of peaks positions between calibrated spectra and standard spectrum is eliminated to less than 3keV in the whole energy region of 0~3MeV.

## 3.3 Results of detection efficiency

### 3.3.1 Experimental results

The efficiency calibration experiment was carried out in the two water tanks mentioned in section 2.1 and the measured spectra are shown in Fig 7. During experiment, the detectors were hung in the center of the tank with cables. To have smaller uncertainties of measurement, each experiment lasted for several days, except for the measurement of $^{234}$Th with 3"×3" $LaBr_3$:Ce detector, whose results were used for self-calibration validation in Section 3.2 as well. Considering the different sizes of the two scintillators, the counts of each spectrum are normalized to the volume of each scintillator for illustration purpose. With the algorithm of SNIP (Sensitive nonlinear iterative peak, Morháč et al., 2009), the background lines can be calculated as the dashed lines in Fig 7, and the peaks of $^{137}$Cs/$^{234}$Th can be fitted with Gaussian functions so that the net count rate can be calculated. According to equation (6), the detection efficiency of the spectrometers for $^{137}Cs/^{234}Th$ can be calculated, as is shown in Table 2.

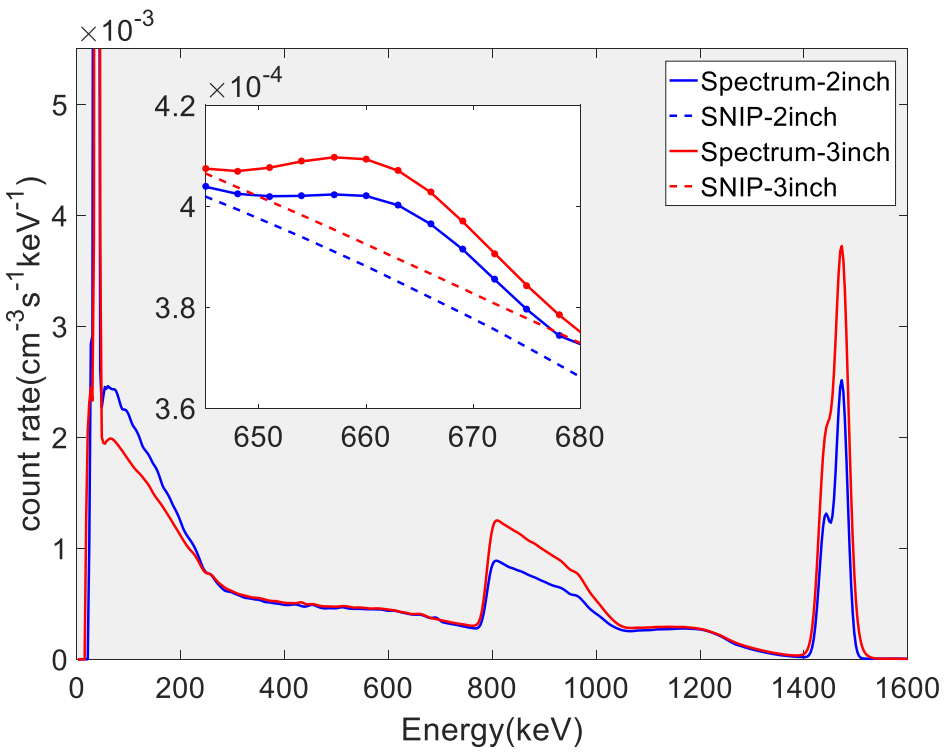

(a) Spectrum of water with $^{137}Cs$

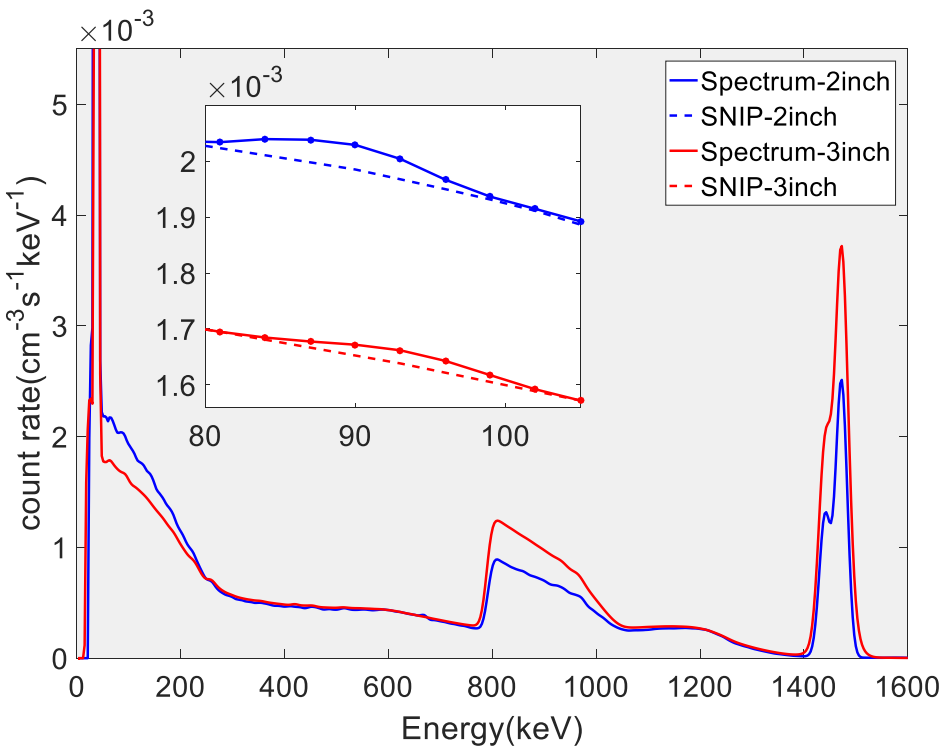

(b) Spectrum of water with $U_3O_8$

Fig 7 Spectrum of 3"×3" $LaBr_3$:Ce detector in water tank. Count rate is normalized to the volume of crystals for better demonstration.

Table 2 Measured efficiency of 3"×3" and 2"×2" $LaBr_3$:Ce detector for $^{137}Cs$ and $^{234}Th$.

| Spectrometer | 3"×3" detector | | 2"×2" detector | |
|---|---|---|---|---|
| Radionuclide | $^{137}Cs$ | $^{234}Th$ | $^{137}Cs$ | $^{234}Th$ |
| Measure time(h) | 46.2 | 1401.8 | 118.8 | 41.1 |
| Activity(Bq/L) | 0.46±0.01 | 5.86±0.69 | 0.45±0.01[a] | 5.86±0.69 |
| Counts of peak(cps) | 0.119±0.006 | 0.076±0.005 | 0.036±0.002 | 0.037±0.011 |
| Measured efficiency(cps/(Bq/L)) | 0.259±0.013 | 0.013±0.002 | 0.080±0.004 | 0.006±0.002 |
| Simulated efficiency(cps/(Bq/L)) | 0.2642±0.0005 | 0.0135±0.0001 | 0.0867±0.0005 | 0.0062±0.0001 |

*a: The activity of $^{137}Cs$ changes because the 2"×2" detector calibration experiment was carried out after one year later.*

### 3.3.2 Simulated results

To evaluate the efficiency of the two spectrometers, MC simulation with GEANT4 code was applied. The simulated geometry is set as the actual geometry, as is shown in Fig 8. As the sources in the same distance to crystal normally provide same contribution to the spectrum, the simulation was carried out in the form of sphere rather than cylinder. To reduce the variance resulting from different contributions of different distances, the simulation was carried out in every 5cm-thick spherical shell, which means radionuclide is distributed uniformly in 5cm-thick spherical shell and the efficiency in each shell will be added together finally according to the weight of each region's volume. The maximum radius of the spherical shell is 100cm, which is large enough for the detectors (Zeng et al., 2015). The simulated efficiency of $^{137}$Cs and $^{234}$Th is shown in Table 2Table 3. Count rate is normalized to the volume of crystals for better demonstration.

Table 2, which are consistent with the measured results. Since the experiments have verified the validity of simulation, simulation could be applied to calculate the efficiency of the system considering the limitation for experiments. The efficiency for common radionuclides in marine environment monitoring was simulated except for $^{137}$Cs and $^{234}$Th, as is shown in Table 3.

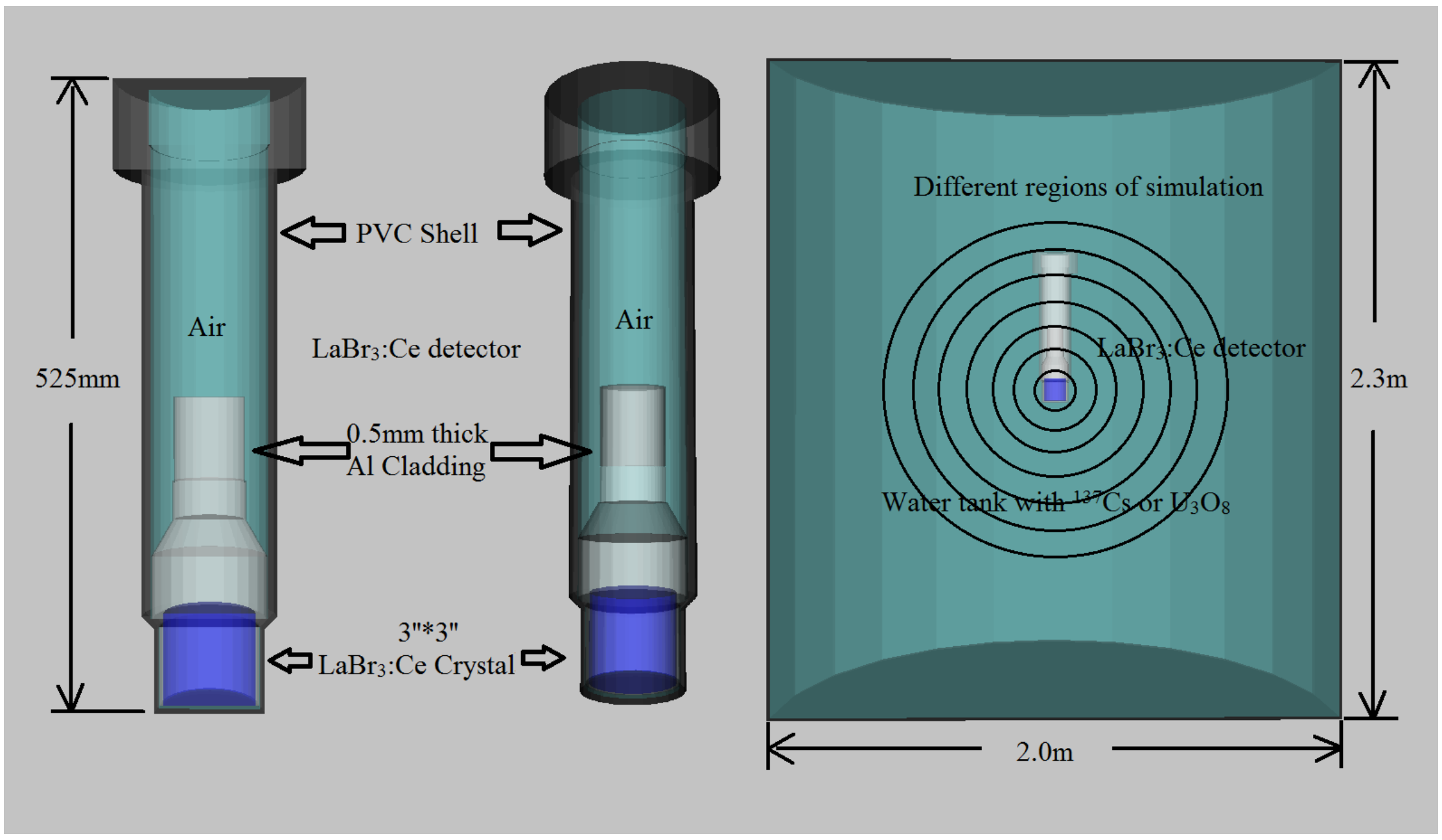

Fig 8 Geometrical model of 3"×3" $LaBr_3$:Ce detector used in the simulation of detection efficiency.

Table 3 Simulated efficiency of 3"×3" and 2"×2" $LaBr_3$:Ce detector for common radionuclides.

| Radionuclide | γ-ray energy(keV) | Emission probability(%) | Efficiency of 3" (cps/(Bq/L) | Efficiency of 2" (cps/(Bq/L) |
|---|---|---|---|---|
| $^{234}$Th | 92~93 | 4.33 | 0.0135±0.0001 | 0.0062±0.0001 |
| $^{131}$I | 364.49 | 81.20 | 0.2868±0.0005 | 0.1113±0.0002 |
| $^{134}$Cs | 569.331 | 15.37 | 0.0492±0.0001 | 0.0165±0.0001 |
| $^{124}$Sb | 602.728 | 97.78 | 0.3112±0.0006 | 0.1013±0.0003 |
| $^{134}$Cs | 604.722 | 97.63 | 0.3119±0.0006 | 0.1005±0.0003 |
| $^{110m}$Ag | 657.75 | 94.38 | 0.2942±0.0005 | 0.0952±0.0006 |
| $^{137}$Cs | 661.659 | 84.99 | 0.2642±0.0005 | 0.0867±0.0005 |
| $^{134}$Cs | 795.868 | 85.47 | 0.2577±0.0005 | 0.0791±0.0005 |
| $^{58}$Co | 810.766 | 99.44 | 0.3008±0.0005 | 0.0924±0.0008 |
| $^{54}$Mn | 834.855 | 99.97 | 0.3002±0.0005 | 0.0932±0.0006 |
| $^{110m}$Ag | 884.67 | 74.00 | 0.2183±0.0004 | 0.0680±0.0007 |
| $^{60}$Co | 1173.24 | 99.85 | 0.2827±0.0005 | 0.0845±0.0004 |
| $^{60}$Co | 1332.508 | 99.98 | 0.2784±0.0005 | 0.0768±0.0003 |
| $^{40}$K | 1460.822 | 10.55 | 0.0301±0.0005 | 0.0079±0.0001 |
| $^{124}$Sb | 1690.984 | 47.46 | 0.1280±0.0002 | 0.0348±0.0003 |
| $^{208}$Tl | 2614.511 | 99.76 | 0.248±0.003 | 0.0590±0.0007 |

The dependency between energy of γ-ray and detection efficiency is shown in Fig 9. The data points in Fig 9 come from Table 2 (experiment) and Table 3 (simulation) but the calculation of detection efficiency takes no account of emission probabilities, namely the emission probabilities are supposed to be 1. The same trend for 3"×3" and 2"×2" $LaBr_3$:Ce detectors is discovered from Fig 9. At low energy region, the efficiency increases as the energy increases because γ-ray with higher energy is less likely to be prevented from entering the sensitive volume by water. However, at high energy region the tendency becomes opposite because γ-ray with higher energy is more likely to escape from the volume without total energy deposition. The maximum efficiency appears at the energy of 250keV and 200keV for the 3"×3" and 2"×2" $LaBr_3$:Ce detector respectively.

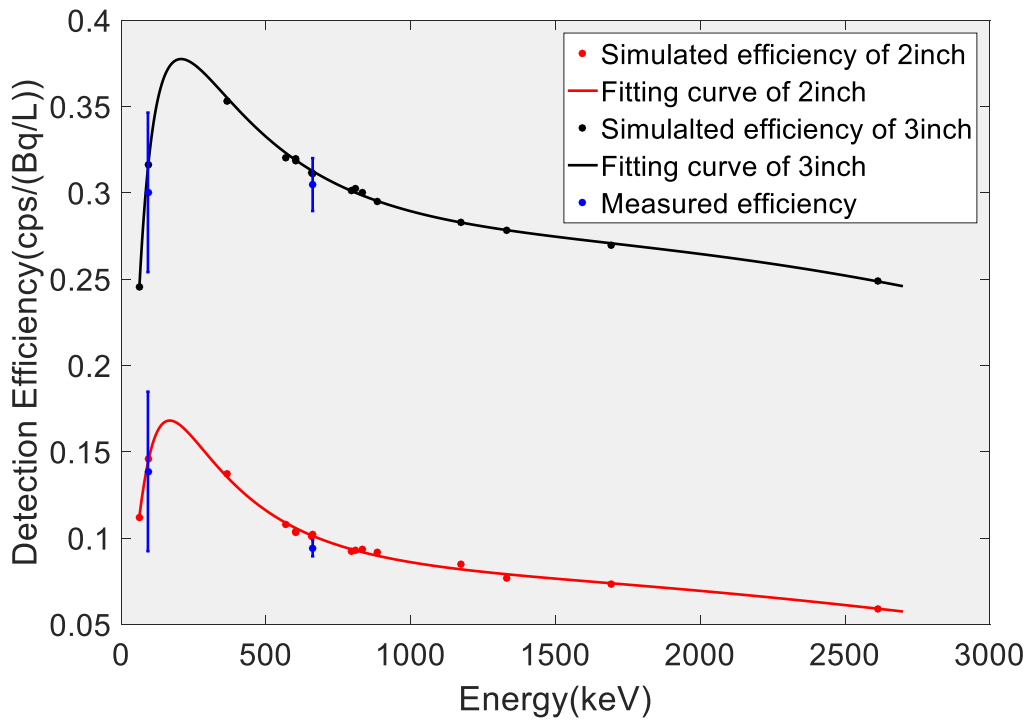

Fig 9 Detection efficiency of the spectrometers.

The dependency between efficiency and energy at higher energy region can be fitted with the following function (Casanovas eta al., 2014):

$$\log \varepsilon_V = \sum_{n=0}^{6} a_n \cdot (\log E)^n \tag{9}$$

Where $\varepsilon_V$ is the volumetric efficiency at the energy E and $a_n$ represents the fitting coefficients.

The curve matches the experimental and simulated results very well and the detection efficiency of the detector at different energy can be achieved with the curve during future actual measurement.

## 4. Conclusion

The recently developed underwater in-situ $LaBr_3$:Ce spectrometer exhibits superior features for marine environment monitoring. To improve the accuracy of measured results with $LaBr_3$:Ce spectrometers, energy calibration and efficiency calibration methods were proposed in this work. With the application of self-calibration by triple peaks from the intrinsic radionuclides, $^{138}$La and $^{215}$Po, in $LaBr_3$:Ce scintillator, the peak-shift caused mainly by change of temperature can be eliminated effectively in the whole energy region of 0~3MeV. Thus, the system's stability can be guaranteed in the region of 0~3MeV during long-term underwater in-situ monitoring.

In addition, the detection efficiency of the spectrometers at different energy in the region of 0~3MeV was calculated by MC simulations. The simulated results matched well with those obtained from measurements in water tanks full of diluted $^{137}$Cs and $U_3O_8$ solution. Different tendencies in high energy level and low energy level were validated by 662keV γ-ray from $^{137}$Cs and 92.5keV γ-ray from $^{234}$Th. Relationship between energy and efficiency was fitted in the whole energy region, which provides an easier way to obtain the efficiency for future measurement.

## Acknowledgments

This work was supported by Public science and technology research funds projects of ocean (No. 201505005); National Key Scientific Instrument and Equipment Development Project (2016YFF0103902); the Ministry of Science and Technology special foundation work (2012FY130200) and Xiamen Southern Ocean Research Center Project (14GZB014NF14)

**Figures**

**Tables**